\documentclass[10pt]{article}
\usepackage[T1]{fontenc}
\usepackage[cp1250]{inputenc}
\usepackage{graphics}
\usepackage{eepic}
\input epsf

\begin{document}
\begin{center}
SIMPLE MODEL FOR THE DYNAMICS OF CORRELATIONS IN THE EVOLUTION OF
ECONOMIC ENTITIES UNDER VARYING ECONOMIC CONDITIONS\\ [5mm]
Marcel AUSLOOS$^1$, Paulette CLIPPE$^2$ and Andrzej P{\c
E}KALSKI$^3$\\ [5mm]

$^1$ SUPRAS and GRASP, B5, Sart Tilman Campus, B-4000 Li$\grave
e$ge, Euroland, \\$^2$ GRASP, B5, Univ. de Li$\grave e$ge,, B-4000
Li$\grave e$ge, Euroland, \\$^3$ Institute of Theoretical
Physics, University of Wroc{\l}aw, pl. Maxa Borna 9, \\PL-50-204
Wroc{\l}aw, Poland
\end{center}

{\it Abstract}

From some observations on economic behaviors, in particular
changing economic conditions with time and space, we develop a
very simple model for the evolution of economic entities within a
geographical type of framework. We raise  a few questions and
attempt to investigate whether some of them can be tackled by our
model.  Several cases of interest are reported. It is found that
the model even in its simple forms can lead to a large variety of
situations, including: delocalization and cycles, but also
pre-chaotic behavior.


\vskip 0.5cm

\section{Introduction}

It is of usual knowledge through personal or through media
information that changing economic world  conditions take place
on different time and space scales. Words like  globalization or
mondialization and  delocalization are used to explain economic
problems everywhere, and often serve as excuses by politicians.
Basic questions  are whether the consequences of such politics
can be predicted, or even can be avoided.

Moreover, for a long time it has been thought that there are so
called economic cycles,\cite{cycleK,cycle2,cycle3} either for the
world economy or for specific goods or companies. In market
economies, the clustering of turns in output, income, employment
and sales reflects a so called  cyclical movement that is the
hallmark of business cycles.

From an econophysicist point of view, using statistical
mechanics  as a primary support of investigations, nature seems
to work as a  "self-organized system", \cite{Bak} characterized
by scaling laws. The models and theories pertaining to these with
punctuated equilibrium features rarely contain cycles. Are these
considerations useful in economy  ? ... interesting question !

Consider a recent specific political event : the Berlin wall
falling  and opening of markets in Eastern Europe and Central Asia
to a so called liberal economy. From a physicist point of view,
one would say that there is an increase in "physical volume, or
available space". This has practically led to so called
delocalization, in the sense that there could be a flow of
entities into the recently opened space. However this feature
also occurred inside the European Union, where there is also a
disparity in "external field conditions", like the tax systems,
or different workers' skills or wages, weather, available
information, ... The external fields, in terms of physics, may
thus be considered to be of legal, economic, political,
meteorological, .. origin.

One could first ask whether one can describe , within a simple
approach, the concentration of enterprises, their spatial
distribution, their so called "fitness" under varying in time and
space economic field conditions.  We do not aim at deriving or
writing  a Hamiltonian or evolution equations here, but rather
would like to simulate time and space-dependent field conditions
and observe whether this leads to a non-trivial behavior.

In Sect. 2, we will present a model, very simple but already
sufficiently elaborate in order to take into account most of
basic economic conditions, - in words.  In section 3, we will
outline the simulation technique used for implementing the model.
We will present a few results in Sect.4, and end with a
conclusion in Sect.5.

\section{Model}

The model has for ingredient a partition of a lattice (the
world!). Here the whole world (or Europe ?) is divided into N (=
3) regions  of  equal size ({\it Gallia  est omnis divisa in
partes tres} \cite{Caesar}). One could imagine a Far East, Middle
East and West division, or a Western Europe, a Central Europe and
an Eastern Europe, or also reduce the scale to three parts in a
specific country, like Belgium. Scales are known to be avoided in
modern statistical physics anyway.

At the beginning all "firms"  (enterprises, agents, ..)  are
located only in the ''West`` (region  I). We suppose that there
is a barrier ("an iron curtain") preventing companies to enter
the ''Central`` (II) and ''East" (III)  regions.  In the
following, there is never any enterprise in the Central nor East
region at the beginning, as long as the "wall" exists. At the
beginning therefore, enterprises in the West, evolve according to
the evolution law described here below, which will have always
the same form also when the wall is destroyed, and for all regions
of the world (so called "globalization").

In the West there was a certain initial number ("concentration")
$c$  of enterprises, though its precise value is not very
important. Each company ($i$) is characterized by one real
number  $f_i \in $\,[0,1],  which could be called a "fitness",
like in many other recent investigations.
\cite{BakSneppen,AMPV,APKSV}

Each region   is characterized by economic conditions, which are
assumed to be an "external field", represented by $one$  real
number,  $F$ also belonging to [0,1]. The best condition, by
symmetry, is  $F = 0.5$

At the beginning the conditions in the West are optimal while in
the Central and Eastern parts they do not matter, since there is
no company belonging to the western scheme. After a certain time
($t_1$) the "Berlin wall" is destroyed, and the regions II and III
are invaded according to a simple diffusion rule.

After another time $t_2$, the value of  $F$ in all regions can be
(and will be) changed. Usually, i.e. for an evolution toward a
globalization, the conditions in the Central and East regions
should be  (after the first change) worse than in the West
region. However each subsequent change makes life easier in the
Center and East, and more difficult in the West. After all it got
the best conditions at the beginning !!!

Questions to be raised, concern whether enterprises survive or not, get better or
worse, whether there is a drastic move toward Central and East, or not.

\section{Monte Carlo Simulation}

We consider a  square lattice. The "horizontal" size of the first
region is $L_x$ = 150. The "vertical size" of each region is $L_y$
= 100. We have only done some simulation for equal size regions.,
i.e. the border between the first and the second region is at
$x_1$ = 50, and between the second and the third one at $x_2$ =
100. The Monte-Carlo (MC) algorithm  describing the dynamics of
the system is the following :

1. At each time step one firm ($i$) is randomly chosen; it's survival probability
is calculated from the formula
\begin{equation}
 p_i = \exp(-sel |f_i - F|),
\end{equation}
 where $sel$ is a parameter describing  a {\it selection pressure},
 in other words how demanding is the environment. This may mean e.g. the amount
of competitors, the availability of goods, the production of goods, .... It is in
the present investigation constant in time and space.

If the enterprise did not survive, because $p_i$ is lower than  a
randomly chosen number $r \in$ [0,1], the program returns to 1.

2. If the $i$ company is fit enough, this  enterprise attempts to
move into any one of the  (4) nearest neighbor sites. If there is
no vacant site, return to 1. However if one empty site is found
the administration board of the company  has a "business plan" :
it is considering two possibilities. If there is another company,
$j$, in the nearest neighborhood, then with a probability  "$b$"
(always = 1 $\% $ here below) the  $i$ enterprise merges with the
second ($j$) one,  and $j$ is eliminated.

The fitness of  $i$ is changed according to the formula

\begin{equation}
f_i = 0.5*(f_i + f_j) + sign[0.5 - r]* |f_i - f_j|)*0.54,
\end{equation}
where  $r$  is a random number  in [0,1]. The choice of the last
factor in eq.(2) equal 0.54 leads to several possible dynamics of
the system. Increasing or lowering it would result in fast
homogeneization of the system.

In the other possibility, thus with a probability $(1-b) \%$, the
two enterprises produce new enterprises or spin-off's;   here
below we consider only the case in which only two enterprises $k$
and $m$ are founded.

However there must be some "room" (vacant sites) in the Moore
neighborhood (8 sites on the square lattice) of the first
enterprise in order to create a new one. If there is less space,
less new firms  (thus one only or no at all) may be  created.

After choosing all agents, one MC time step is done. We have
restricted the simulation time in each run to 5000 MCS, since
afterwards we observed rather small oscillations around the
stationary state.

\section{ Results}

There are several ways of presenting the results.  In the
following we stress cases demonstrating the pertinence of the
model and giving arguments for the coincidence with the numerical
results of qualitative observations .

We report  three cases ($N, A, B$), differing by the sequence of
changes of the external field values in the three regions. In the
case $N$ the field values (selection pressure) in the first region
do not change, whereas after 100 MCS, i.e. fall of the "iron
curtain", the selection in the second and third region read, 0.4
and 0.3, respectively. In the case $A$ after $t_1$ = 100 MCS,
which is again the only change, the selection in the I, II and III
regions is 0.40,0.50 and 0.60, respectively. Finally, in the case
$B$ there are two changes. One at 100 MCS, after which the
conditions in the three regions are 0.40,0.50,0.60, and after the
second change at $t_2$ = 300 MCS  the values of the external field
are 0.45,0.50 and 0.55, in the I, II and III regions.

\begin{figure}
\centerline{\epsfxsize=6cm
\epsfbox{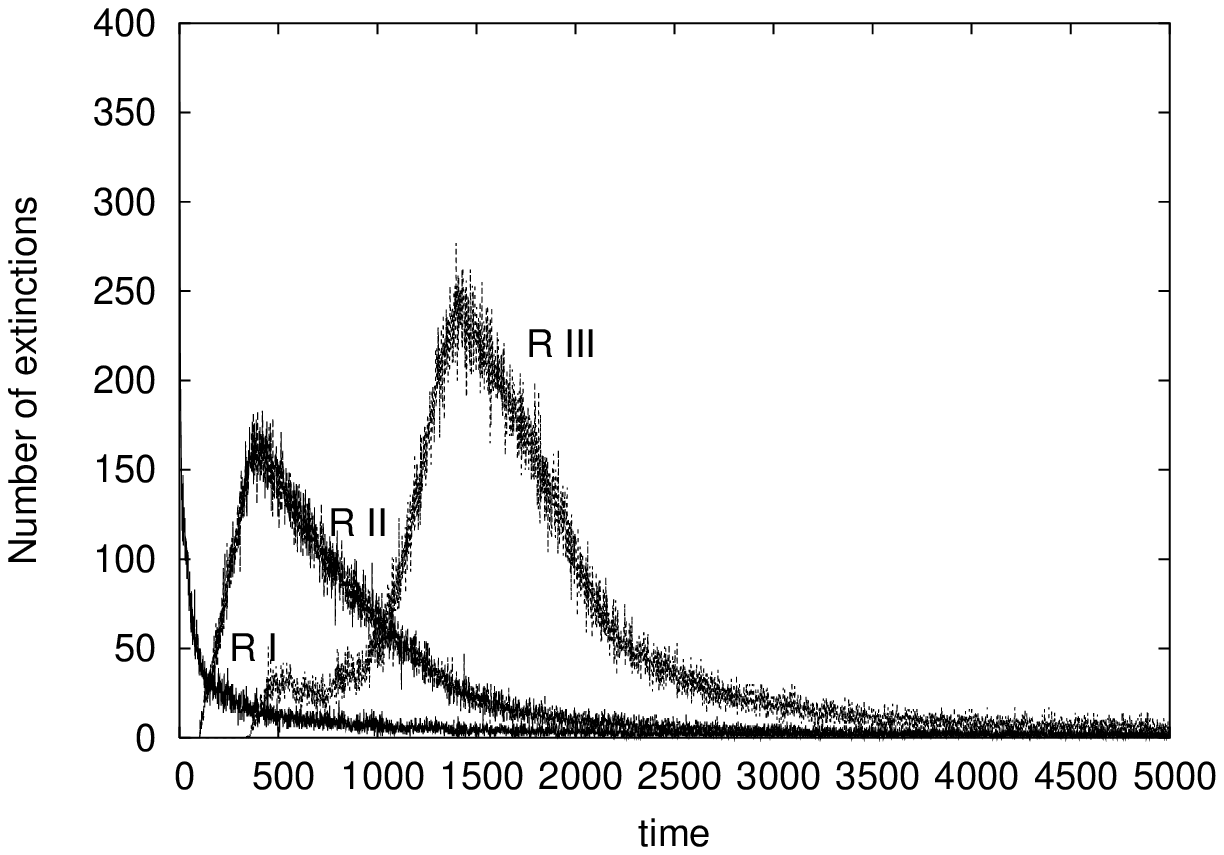}\epsfxsize=6cm\epsfbox{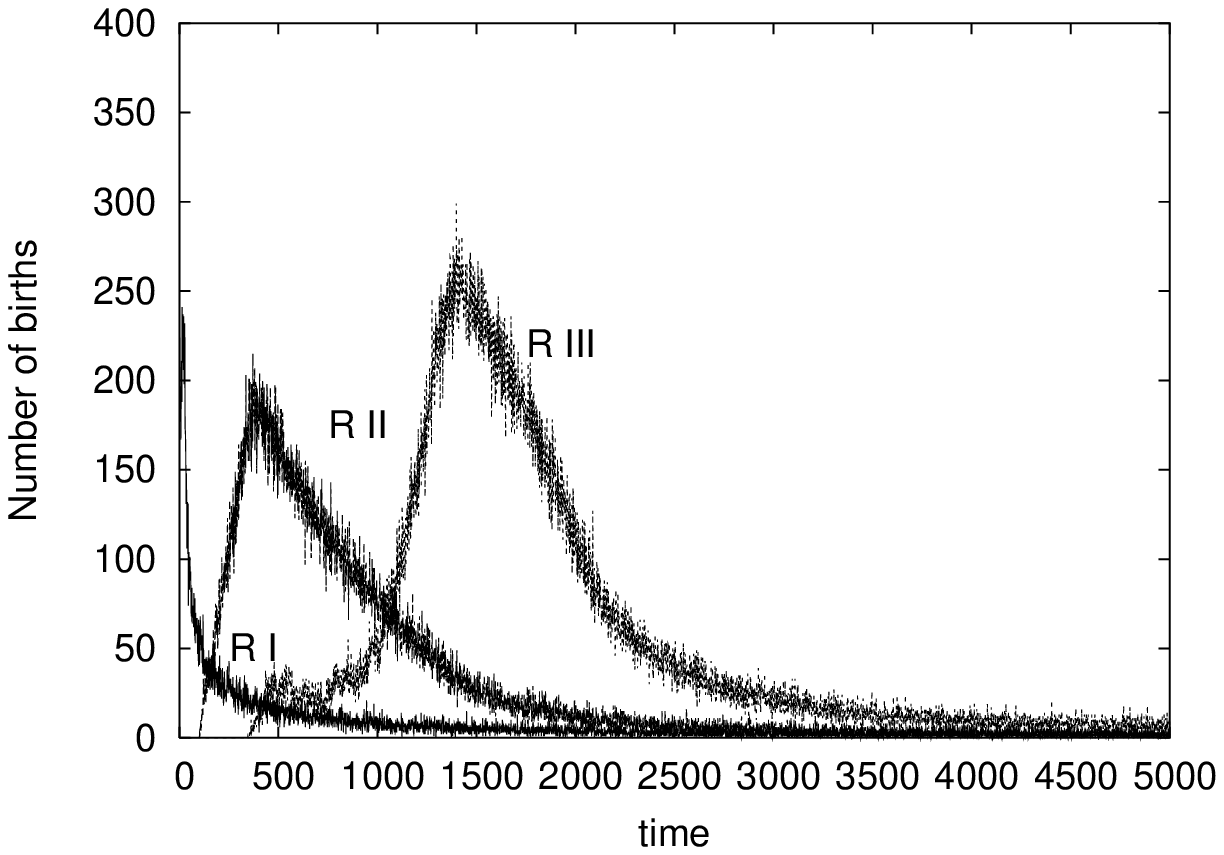}}
\caption{ Case $N$, with external field values $F$ =0.5, 0.4, 0.3
in the three regions after $t_1 = 100 MCS$; (a) number of
enterprise extinction and (b) births as a function of time (in
MCS)  in the three regions for a $sel$ =1.0}
\end{figure}

 In each case, we start with an initial concentration $c$ =0.5
in region I, and destroy ''the Berlin wall``  after 100 MCS.
Starting with case $N$,  we show in Fig. 1(a-b) the number of
enterprise extinction and births as a function of time in the
three regions, for a weak selection, $sel$ =1.0. The fitness and
concentration in the three regions at different times are shown
in Fig.2(a-b). There is a rapid increase in concentration in the
opened region as soon as it is open, but the death and birth
process is like an equilibrium one.  The fitness optimal value in
each region is easily reached and remain stable in time.

\begin{figure}
\centerline{\epsfxsize=6cm \epsfbox{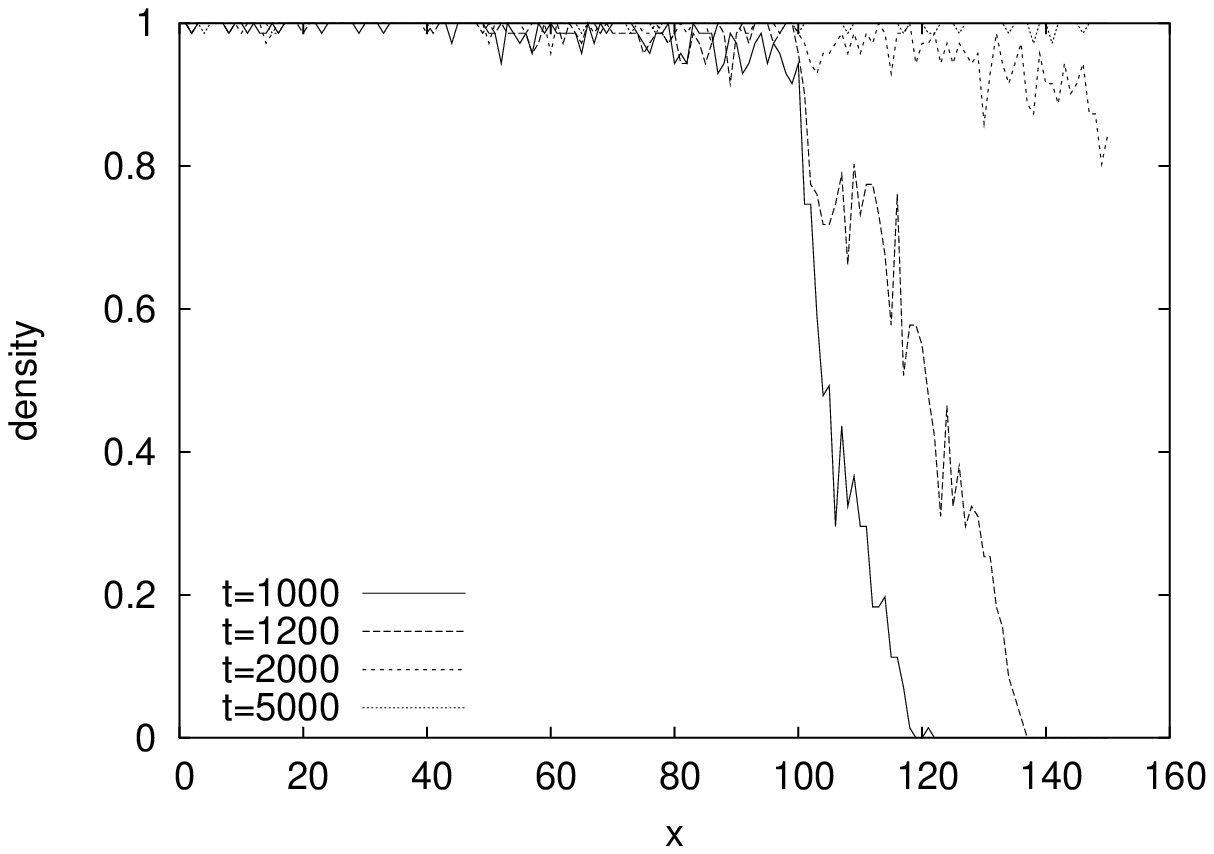} \epsfxsize=6cm
\epsfbox{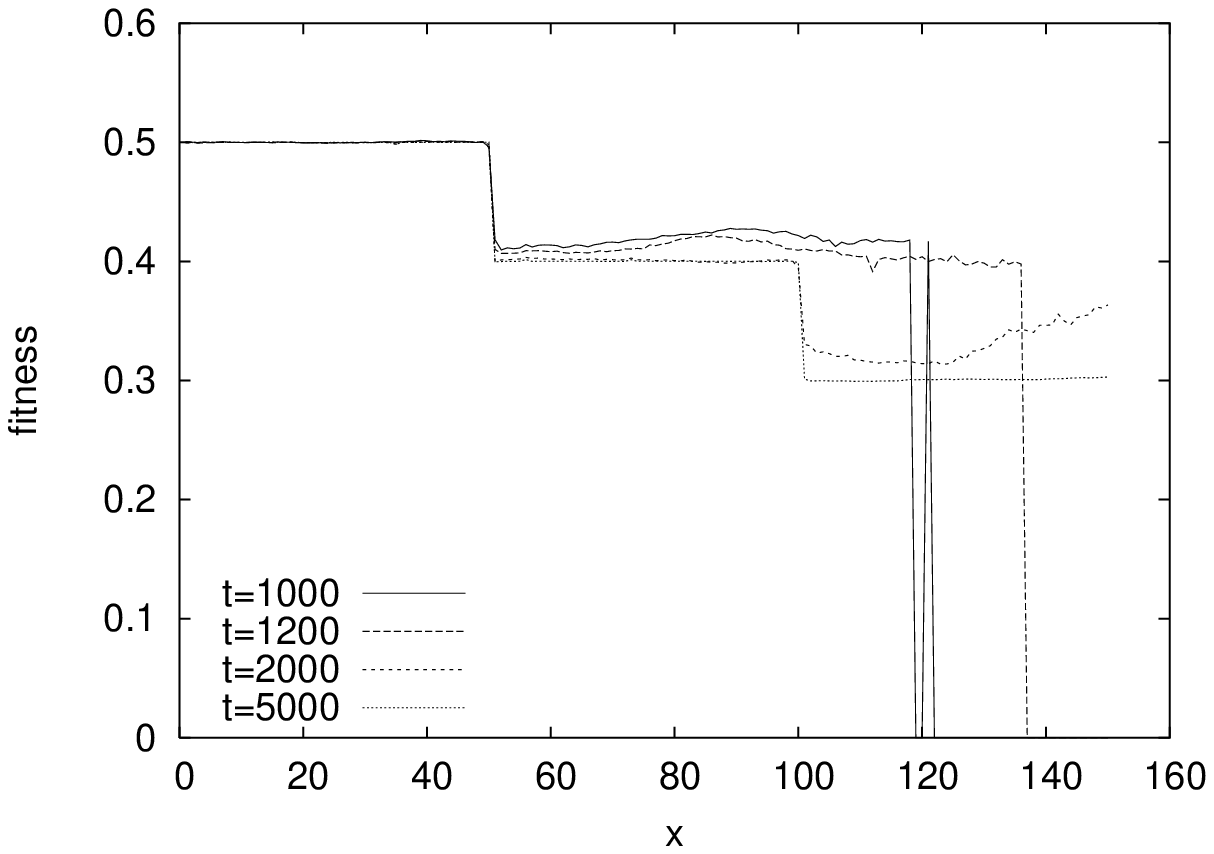}} \caption{ The (a) concentration and (b)
fitness in the three regions taken at different times in case $N$,
i.e. with external field values $F$ =0.5, 0.4, 0.3 for $sel$ =
1.0}
\end{figure}

The same type of figure as Fig.2 is shown in Fig.3(a,b)  but for a
higher selection pressure, $sel$=1.5. One observes a greater
invasion difficulty process due to the higher $sel$ value, i.e.
for a more demanding environment.

\begin{figure}
\centerline{\epsfxsize=6cm \epsfbox{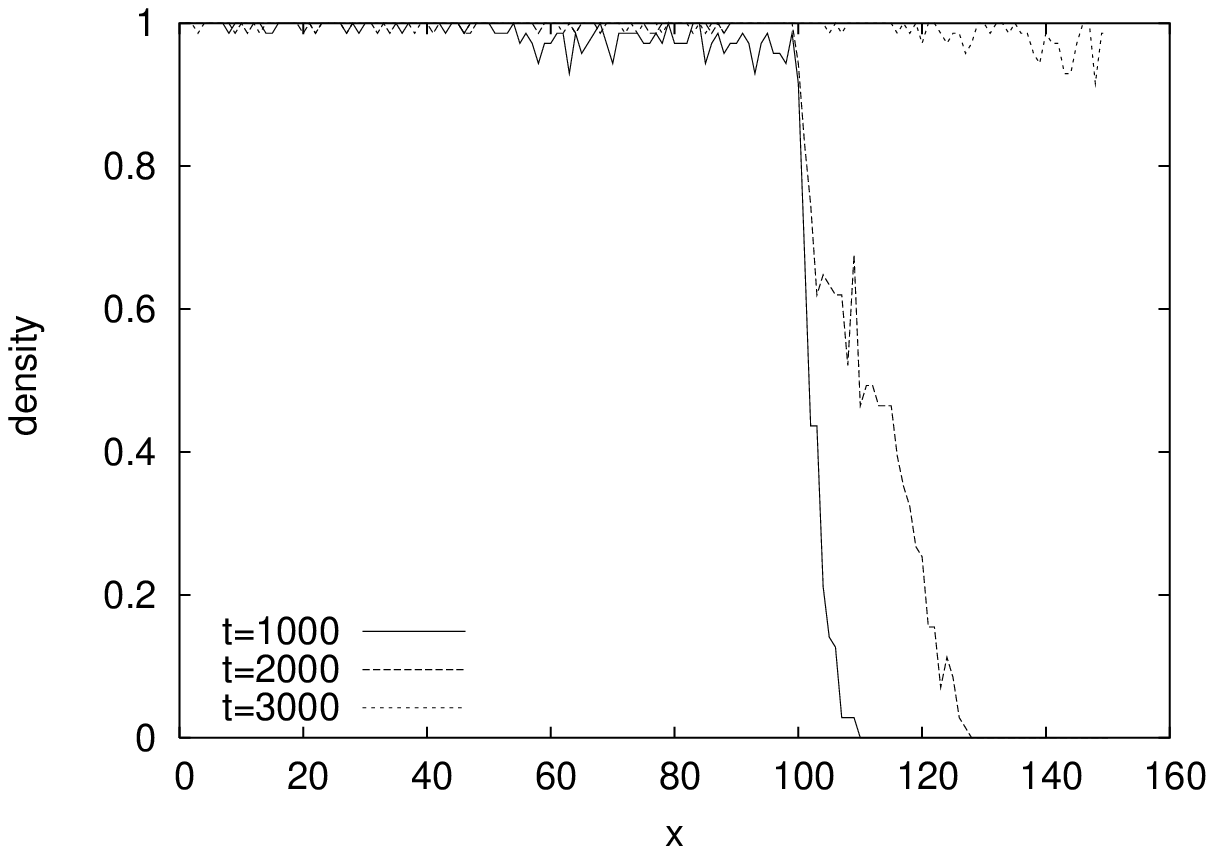} \epsfxsize=6cm
\epsfbox{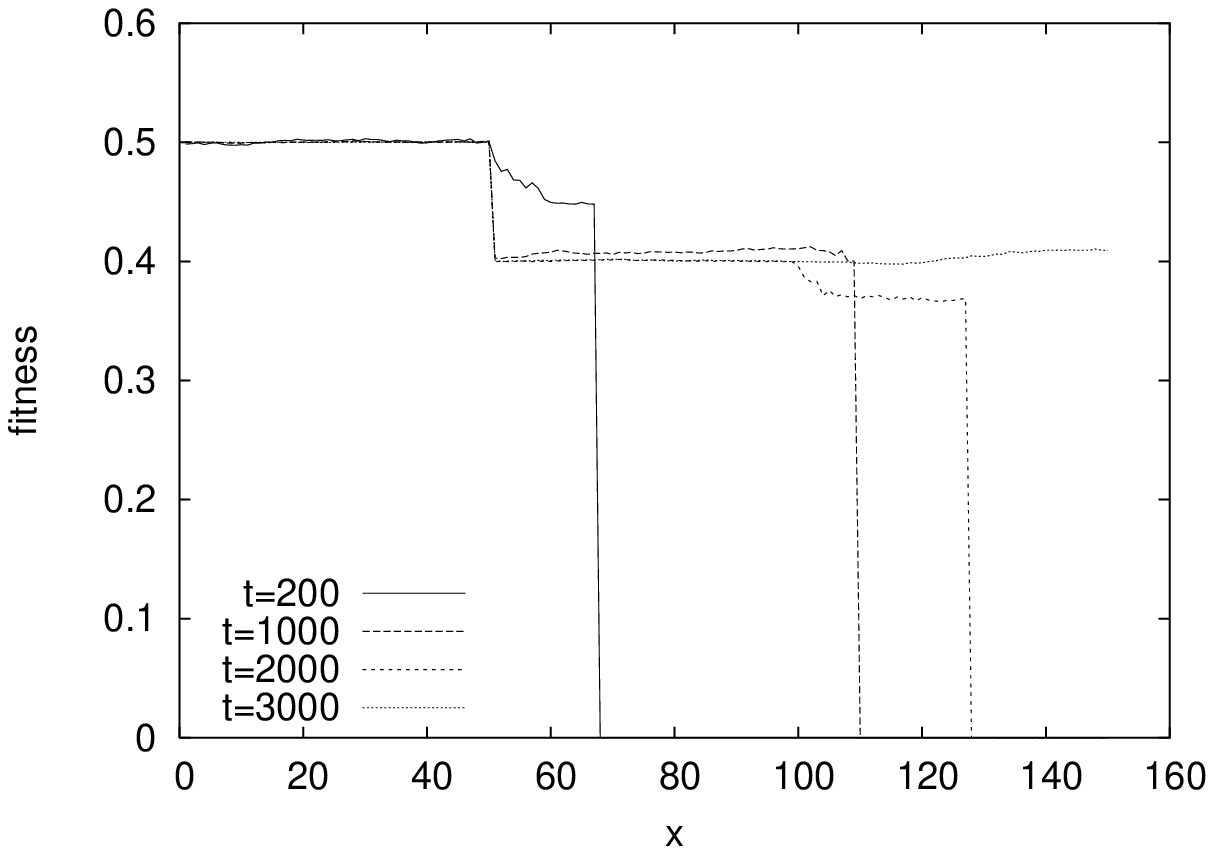}} \caption{ Same as in figure 2, but for a
stronger selection, $sel$ = 1.5}
\end{figure}

In case $A$, we consider that after the opening of ''the iron
curtain``, the optimal situation is found in region II; we  impose
field conditions  in II and III to be  0.4 and 0.6 respectively.
In Fig. 4(a-b) we show the concentration in the three regions
taken after different number of MCS, but for a still larger (than
in previous figures) $sel$, i.e. $sel$= 1.75 in Fig. 4a and 2.75
in Fig. 4b. This latter case is remarkable since, after 1000 MCS
the first region becomes empty. This is because the second region,
under a 0.5 field, has better conditions. The concentration of
enterprises in II hereby increases up to its maximum values.

\begin{figure}
\centerline{\epsfxsize=6cm \epsfbox{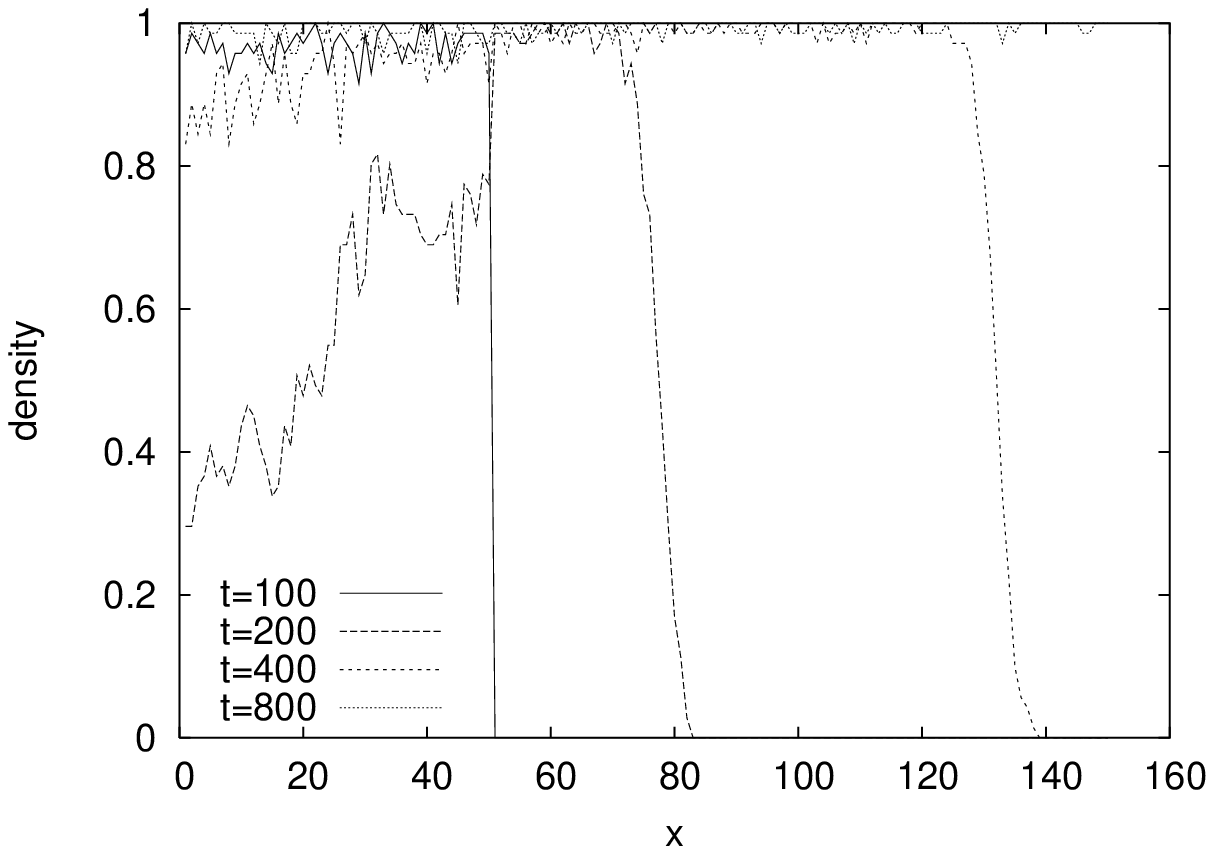} \epsfxsize=6cm
\epsfbox{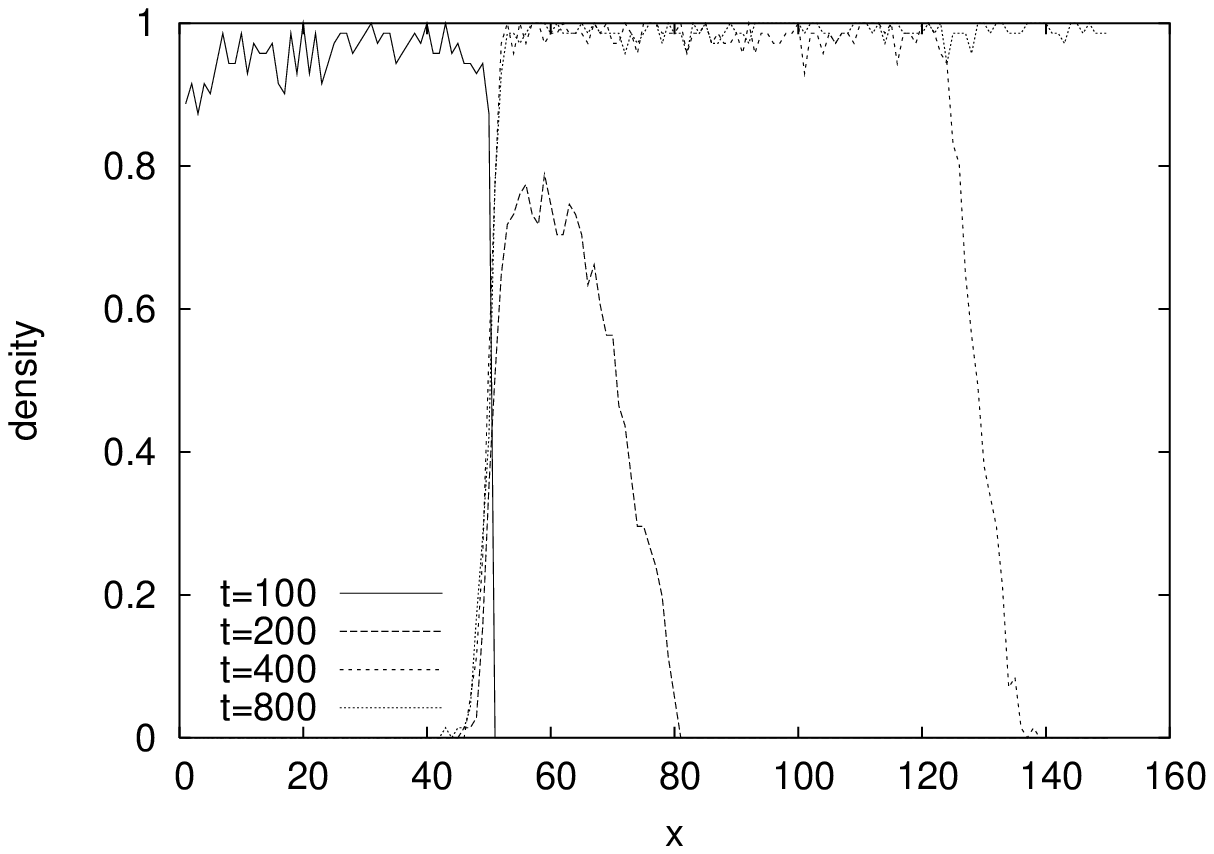}} \caption{ The concentration in the three
regions taken at different times  in case A, i.e. with external
field values $F$ =0.4, 0.5, 0.6 for (a) $sel$ = 1.75 and (b)
$sel$=2.75}
\end{figure}

However with increasing time, the third region is invaded as well.
One might wonder whether by symmetry the first region could be
re-invaded as well. The answer is  that depending on the initial
distribution of agents, here determined by the seed of the random
number generator, two cases can occur: either region II and III
are filled up, with enterprises reaching their appropriate
optimal fitness, or region I on one hand, and regions II and III
on the other hand, can also be filled up, with enterprises at
their optimal fitness. This indicates that the model contains
some bifurcation point at some value of the selection pressure,
here it happened for $sel$ of the order of 1.8.

 The invasion of the third region is illustrated in
Fig.5a,  for model $B$ and $sel$ = 2.6,  showing the density of
the enterprises at the three regions at the early (a) and later
(b) stages of the evolution. The external conditions in the three
regions after 300 MCS have the following values
  0.45, 0.50 and 0.55 in the I, II and III regions resp.  It is
 observed that the first
region  is depleted in favor of region II, but thereafter the
first one is  replenished again. The beginning of a cycle is
therefore found.

\begin{figure}[here]
 \centerline{\epsfxsize=6cm \epsfbox{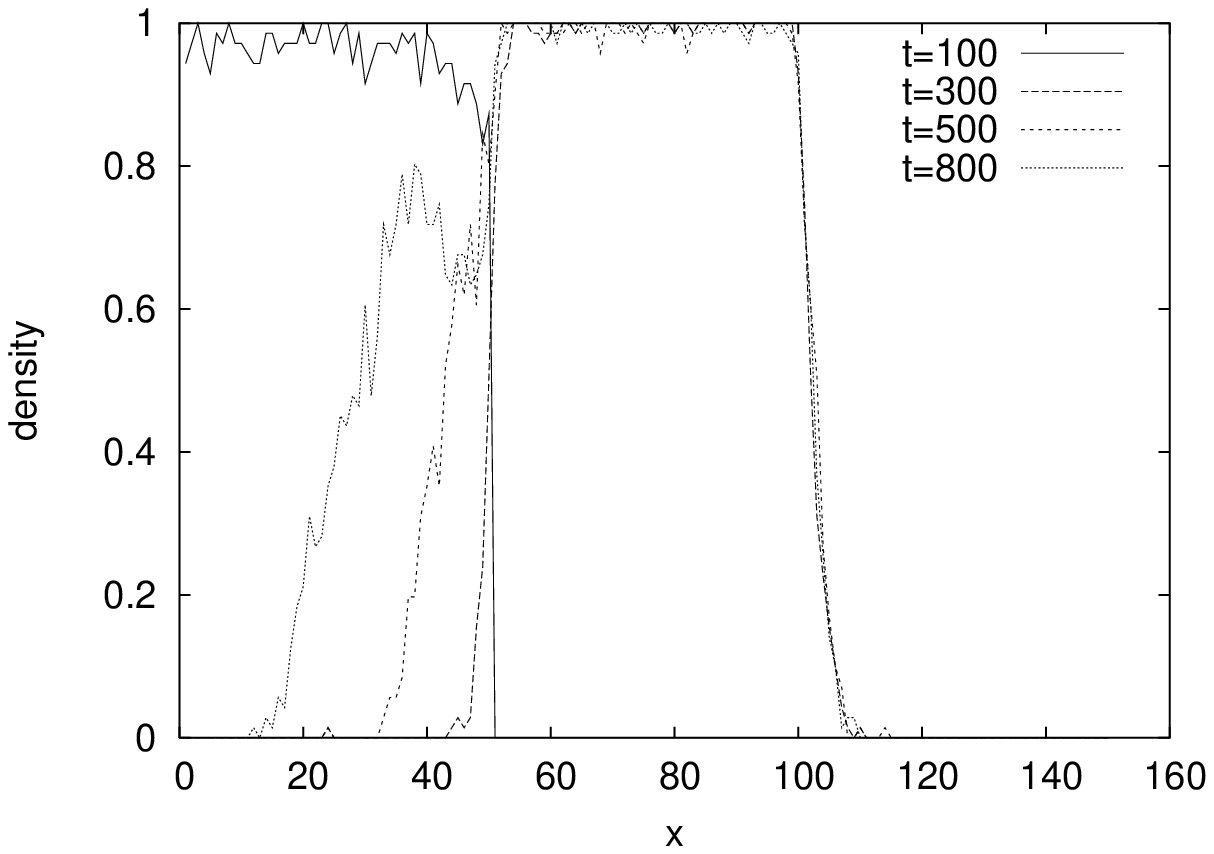}\epsfxsize=6cm
 \epsfbox{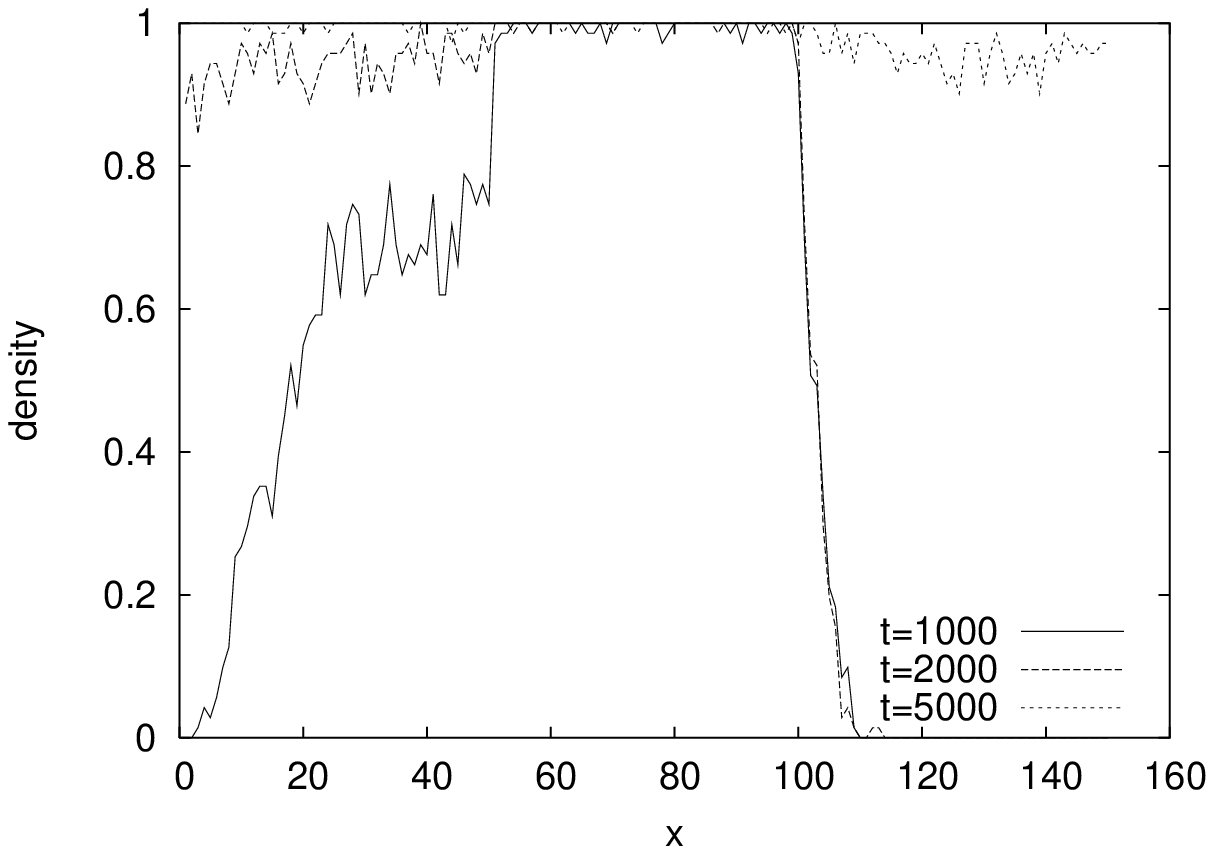}}
\caption{ The concentration in the three regions taken at
different times in case B, i.e. with external field values $F$ =
0.4, 0.5, 0.6 after 100 MCS and $F$ = 0.45, 0.50, 0.55 after 300
MCS. {\it sel} = 1.75.(a) shows earlier stages and (b)  later
stages of evolution}
\end{figure}

\section{Conclusions}

From this set of  results,  we observe that the model contains
many facets : (i) there are relatively well marked effects due to
the "selection pressure"; (ii) an adaptative economic
("external") field time sequence can imply stable  or unstable
density distributions; (iii) the diffusion process rule(s) are
useful for invasion process, but are also relevant for
replenishing abandoned regions; (iv) the diffusion rules together
with the business plan, and the selection pressure seem to lead
to a quasi equilibrium state with respect to birth and death
processes; (v) interesting sort of cycling and bifurcation
features have been observed, which might indicate either complex
oscillations or chaotic behaviors. \cite{Addison}

We are aware that further improvements  are needed. We are
drastically caricaturing macro and micro economy field
conditions, as well as the description of the "internal"
interactions sequence(s). Moreover a company should not be
described by one scalar number $f_i$, but rather a vector model
coupled to a (so called external) vector  field should be
examined. Moreover the birth and death  process description
through merging and spin off's is also to be improved. Analytical
work could be of interest to describe bifurcation conditions, and
implement different company size distributions in different
countries as in ref.[10]

\vskip 0.6cm

{\bf Acknowledgments}

\vskip 0.6cm

MA and PC thank the organizers of the Bali Econophysics conference very much for
their welcome, including grants, stimulating discussions and comments. PC thanks
also FNRS for financial support.\vskip 1cm

%\vskip 2.6cm
\noindent
(*$^1$) email:   marcel.ausloos@ulg.ac.be \\ ($^2$)
email:   P.
Clippe@ulg.ac.be \\ ($^3$) email:   apekal@ift.uni.wroc.pl \\


\begin{thebibliography}{99}

\bibitem{cycleK} M. Kalecki, "A Macroeconomic Theory of the Business Cycle", {\it
Econometrica} {\bf 3},  327-344 (1935);  ibid., "A Theory of the Business Cycle",
{\it Review of Economic Studies} {\bf 4},  77-97  (1937); ibid., "Theory of
Economic Dynamics: An essay on cyclical and long-run changes in capitalist
economy", (Monthly Review Press, New York, 1965)

\bibitem{cycle2} J.B Long and  C.I Plosser, "Real Business Cycles", {\it J.
Polit. Econ.} {\bf  91},  39-69  (1983)

\bibitem{cycle3} S. Basu and A.M. Taylor, "Business cycles in international
historical perspective", {\it J. Econom. Perspect.} {\bf  13}, 45-68  (1999)

\bibitem{Bak}  P. Bak, "How Nature Works: the science of self-organized
criticality", (Oxford UP, Oxford, 1997)

\bibitem{Caesar}  J.  Caesar, "De Bello Galico"; S. A. Hanford and J.F. Gardner,
"The Conquest of Gaul", (Penguin Books, London, 1982)

\bibitem{BakSneppen} P. Bak and K. Sneppen, "Punctuated equilibrium and
criticality in a simple model of evolution", {\it Phys. Rev. Lett.} {\bf 71},
4083-86 (1993)

\bibitem{AMPV}  M. Ausloos, I. Mroz, A. P{\c e}kalski, and N. Vandewalle, "Lattice gas
model of gradual evolution", {\it Physica A} {\bf  248},  155-164 (1998)

\bibitem{APKSV}  A. P{\c e}kalski and K. Sznajd-Weron, "Population dynamics with and
without selection", {\it Phys. Rev. E} {\bf  63},     31903 (2001)

\bibitem{Addison} P.S. Addison, {\it Fractals and Chaos}, (Institute of Physics,
Bristol, 1997)

\bibitem{Ramsden} J.J. Ramsden and Gy. Kiss-Hayp\'al, "Company size distribution
in different countries", {\it  Physica A} {\bf 277},  220-227  (2000)

\end{thebibliography}
\end{document}